\documentstyle[12pt]{article}
\font\cmssB=cmss17 scaled \magstep1

\font\tenrm=cmr10 scaled\magstep1
\font\teni=cmmi10 scaled\magstep1
\font\tensy=cmsy10 scaled\magstep1
\font\tenex=cmex10 scaled\magstep1
\font\tenbf=cmbx10 scaled\magstep1
\font\tenit=cmti10 scaled\magstep1

\font\sixrm=cmr6 scaled \magstep1
\font\sixi=cmmi6 scaled\magstep1
\font\sixsy=cmsy6 scaled\magstep1
\font\sc=cmr5 scaled\magstep1
\font\ssmall=cmr5 scaled \magstep1

\font\fivemi=cmmi5 scaled\magstep1
\font\fivesy=cmsy5 scaled\magstep1
\font\small=cmr8 scaled \magstep1
\font\smmath=cmmi8 scaled\magstep1
\font\sit=cmti8 scaled\magstep1
\font\smsy=cmsy8 scaled\magstep1
\font\sbf=cmbx8 scaled\magstep1
\font\smex=cmex8 scaled\magstep1
\font\smallr=cmr5
\font\smalli=cmmi5
\font\smallsy=cmsy5
\def\rm{\fam0 \tenrm}  
\def\bf{\tenbf}  \def\it{\tenit} 
\rm

\def\footnoterule{\kern-3pt
                  \hrule width 2truein \kern 2.6pt
                  \vskip5pt}
\newinsert\footins
\def\footnote#1{\let\asf=\empty
  \ifhmode\edef\asf{\spacefactor=\the\spacefactor}\/\fi
  #1\asf\vfootnote{#1}}
\def\vfootnote#1{\insert\footins\bgroup
  \interlinepenalty=\interfootnotelinepenalty
  \splittopskip=\ht\strutbox
  \splitmaxdepth=\dp\strutbox
  \leftskip=0pt \rightskip=0pt \spaceskip=0pt \xspaceskip=0pt
  \textindent{#1}\footstrut\futurelet\next\fo@t}
\def\fo@t{\ifcat\bgroup\noexpand\next \let\next\f@@t
  \else\let\next\fot\fi \next}
\def\f@@t{\bgroup\aftergroup\afoot\let\next}
\def\fot#1{#1\afoot}
\def\afoot{\strut\egroup}
\def\footstrut{\vbox to\splittopskip{}}
\skip\footins=20pt 
\count\footins=1000
\dimen\footins=8in
\def\footmacro(#1,#2){\footnote{$^{#1}$}{\vtop{\parindent=10pt
        \baselineskip16pt
        \hsize=15truecm
        \small 
        \textfont0=\small \textfont\bffam=\sbf \textfont\itfam=\sit
        \textfont1=\smmath \textfont2=\smsy \textfont3=\smex      
        \scriptfont0=\ssmall \scriptfont1=\fivemi \scriptfont2=\fivesy
        \scriptscriptfont0=\smallr \scriptscriptfont1=\smalli
        \scriptscriptfont2=\smallsy #2}}}
\def\vfoot(#1,#2){
  \footmacro(,{\hskip-25pt$^{#1}$\hskip7pt\vtop{\noindent #2}})}
\def\foot(#1,#2){$^{#1}$\vfoot(#1,#2)}

\def\eqno(#1,#2){(#1.#2)}
\def\thebibliography#1{\leftline{\bf References}
 \list
 {[\arabic{enumi}]}{\settowidth\labelwidth{[#1]}\leftmargin\labelwidth
 \advance\leftmargin\labelsep \itemsep 0pt \parsep 0pt
 \setlength{\textwidth}{16cm}
 \usecounter{enumi}}
 \def\newblock{\hskip .11em plus .33em minus -.07em}
 \sloppy
 \sfcode`\.=1000\relax}

\def\mailAbe{E-mail: abe@kurims.kyoto-u.ac.jp}
\def\mailNN{E-mail: nbr-nakanishi@msn.com}

\def\today{\ifcase\month\or
  January\or February\or March\or April\or May\or June\or
  July\or August\or September\or October\or November\or December\fi
  \ \ \number\year}
\begin{document}
\pagestyle{plain}
\thispagestyle{empty} 
\setlength{\oddsidemargin}{.5truecm}
\setlength{\textheight}{23.cm}  
\setlength{\textwidth}{16.cm}
 \vsize=23cm \hsize=16cm
\setlength{\topmargin}{-.5cm}
\setlength{\baselineskip}{19.pt} 
\setlength{\parindent}{25pt}
\setlength{\footnotesep}{10pt}
\setlength{\abovedisplayskip}{15pt}
\setlength{\belowdisplayskip}{15pt}
\linepenalty=1000
\setlength{\overfullrule}{0pt}
\input mssymb.tex
%
\centerline{\cmssB Note Added to ``Proof of the Gauge Independence}
\vskip10pt
\centerline{\cmssB of the Conformal Anomaly of Bosonic String}
\vskip10pt
\centerline{\cmssB in the Sense of Kraemmer and Rebhan''}
\vskip50pt
\centerline{
 Mitsuo Abe$^{\hbox{\sc a},\,}$\foot(1,{\mailAbe})
 and 
 Noboru Nakanishi$^{\hbox{\sc b},\,}$\foot(2,{Professor Emeritus of Kyoto 
                                              University.  \mailNN})}
\vskip15pt
\centerline{\sit $^{\hbox{\sc a}}$Research Institute for Mathematical 
Sciences, Kyoto University, Kyoto 606-8502, Japan}
\vskip0pt
\centerline{\sit $^{\hbox{\sc b}}$12-20 Asahigaoka-cho, 
Hirakata 573-0026, Japan}
\vskip30pt
%
%
\centerline{\bf Abstract}
A comment is given to the reply of Kraemmer and Rebhan to our paper.
\vskip30pt
%
%
In our previous paper\cite{AN}, it has been pointed out that the Kraemmer 
and Rebhan's proof of gauge independence of conformal anomaly of bosonic 
string for various gauge-fixings\cite{KR} is wrong, but the gauge 
independence is proved for the gauge-fixings which reduced to the linearized 
de Donder gauge in the flat limit of the background metric.
After our paper was circulated, Kraemmer and Rebhan have written 
a reply to it\cite{KR2}.  
While they agree to our pointing-out of the noncommutativity
between the BRS transformation and $\delta/\delta\hat g_{\mu\nu}$,
where $\hat g_{\mu\nu}$ denotes the classical background metric,
they still claim the validity of their proof. 
So, in order to avoid misunderstanding, we supplement our claim that their
proof is wrong.
\par
Their claim seems to be based on their wrong belief that the effective action 
contains the conformal anomaly {\it a priori.}
One should recognize that the conformal anomaly {\it cannot\/} be seen in the 
effective action itself.  In order to obtain the conformal anomaly, one must 
take $\delta/\delta\hat g_{\mu\nu}$.
The error committed
so far is that this procedure would be made merely to visualize the already
existing conformal anomaly.  This is {\it not\/} the case, however.  
The truth is that the conformal anomaly term --- more precisely speaking, 
the nonlocal term from which people wish to deduce the prefactor of Polyakov's 
induced action --- is {\it produced\/} by this procedure.  
The fact that the anomaly term is {\it produced\/} is more clearly seen in the 
FP-ghost number current: The FP-ghost number conservation is {\it not\/} 
violated at all in the complete exact solution, but its anomaly is encountered
if one considers the two-point function through $\delta/\delta\hat g_{\mu\nu}$
(see Ref.4). 
\par
The fact that the effective action itself does {\it not\/} have any anomaly 
term independent of $D$ can be shown by applying the Kraemmer-Rebhan gauge 
variation procedure \eqno(2,26) until one {\it totally eliminates\/} 
the contribution from the gauge-fixing plus FP-ghost Lagrangian.\foot(3,{It 
should be noted that there is no reason to reject to consider this limit in
their procedure because no normalization condition is imposed on 
the gauge-fixing plus FP-ghost Lagrangian in their proof.})
It is quite unreasonable to suppose that this limit is nonexistent 
because the anomaly term is a {\it simple sum\/} of the contribution 
(proportional to $D$) from the string Lagrangian and the one (independent of 
$D$) from the gauge-fixing plus FP-ghost one.  
Since this procedure is thoroughly carried out without making any disturbance 
on the BRS exactness, there is no reason to believe the sudden appearance of
$-26$ in the effective action in the case in which the gauge-fixing plus 
FP-ghost Lagrangian is present.  {\it Nonzero anomaly contribution can arise 
only as a consequence of the violation of BRS exactness caused by taking 
$\delta/\delta\hat g_{\mu\nu}$.}
\par
Anyway, {\it Kraemmer and Rebhan never demonstrated the existence of 
the conformal anomaly proportional to $D-26$ without using 
$\delta/\delta\hat g_{\mu\nu}$
in the model considered.}
Thus there is a big logical gap between the gauge invariance of the effective
action itself and that of the conformal anomaly proportional to $D-26$.
If they wish to claim the validity of their proof, they must {\it directly\/} 
prove their belief that the conformal anomaly is contained in the effective
action.

\vskip25pt
%
%

\end{document}